\documentclass[10pt,conference]{IEEEtran}
\IEEEoverridecommandlockouts
\usepackage{cite}
\usepackage{amsmath,amssymb,amsfonts}
\usepackage{algorithm}
\usepackage{algorithmic}
\usepackage{graphicx}
\usepackage{textcomp}
\usepackage{xcolor}
\usepackage{amsmath} 
\usepackage{amssymb}
\usepackage{longtable}
\usepackage{tcolorbox}
\usepackage{multirow}
\usepackage[normalem]{ulem}
\usepackage{url}
\usepackage{xcolor}
\usepackage{ifthen}
\usepackage[normalem]{ulem}
\usepackage{xcolor}

\newboolean{showedits}
\setboolean{showedits}{true} 
\ifthenelse{\boolean{showedits}}
{
	\newcommand{\del}[1]{\textcolor{red}{\sout{#1}}} 
}{
	\newcommand{\del}[1]{} 
	
}

\newboolean{showcomments}
\setboolean{showcomments}{true} 
\newcommand{\id}[1]{$-$Id: scgPaper.tex 32478 2010-04-29 09:11:32Z oscar $-$}

\ifthenelse{\boolean{showcomments}}
{\newcommand{\nbc}[3]{
		{\colorbox{#3}{\bfseries\sffamily\scriptsize\textcolor{white}{#1}}}
		{\textcolor{#3}{$\blacktriangleright$#2$\blacktriangleleft$}}}
	}
{\newcommand{\nbc}[3]{}
	\renewcommand{\del}[1]{} 
	}

\definecolor{ibcolor}{rgb}{1.0,0.2,.4}
\definecolor{dsrcolor}{rgb}{0.5,0.6,0}
\definecolor{cfcolor}{rgb}{0,0.5,0.9}
\definecolor{oldcolor}{rgb}{0.2,0.2,0.2}
\definecolor{tdcolor}{rgb}{1.0,0,0}
\definecolor{oldcolor}{rgb}{0.5,0.5,0.5}
\definecolor{lycolor}{rgb}{0.3,0.3,0.8}

\def\BibTeX{{\rm B\kern-.05em{\sc i\kern-.025em b}\kern-.08em
    T\kern-.1667em\lower.7ex\hbox{E}\kern-.125emX}}
\begin{document}

\title{LEAP: Efficient and Automated Test Method for NLP Software}

\author{\IEEEauthorblockN{Mingxuan Xiao}
\IEEEauthorblockA{\textit{College of Computer and Information} \\
\textit{Hohai University}\\
Nanjing, China \\
xiaomx@hhu.edu.cn}
\and
\IEEEauthorblockN{Yan Xiao}
\IEEEauthorblockA{\textit{School of Cyber Science and Technology} \\
\textit{Shenzhen Campus of Sun Yat-sen University}\\
Shenzhen, China \\
xiaoy367@mail.sysu.edu.cn}
\and
\IEEEauthorblockN{Hai Dong}
\IEEEauthorblockA{\textit{School of Computing Technologies} \\
\textit{RMIT University}\\
Melbourne, Australia \\
hai.dong@rmit.edu.au}
\and
\IEEEauthorblockN{Shunhui Ji}
\IEEEauthorblockA{\textit{College of Computer and Information} \\
\textit{Hohai University}\\
Nanjing, China \\
shunhuiji@hhu.edu.cn}
\and
\IEEEauthorblockN{Pengcheng Zhang\textsuperscript{*}}
\IEEEauthorblockA{\textit{College of Computer and Information} \\
\textit{Hohai University}\\
Nanjing, China \\
pchzhang@hhu.edu.cn}
\thanks{*Corresponding author.}
}

\maketitle

\begin{abstract}
The widespread adoption of DNNs in NLP software has highlighted the need for robustness. Researchers proposed various automatic testing techniques for adversarial test cases. However, existing methods suffer from two limitations: weak error-discovering capabilities, with success rates ranging from 0\% to 24.6\% for BERT-based NLP software, and time inefficiency, taking 177.8s to 205.28s per test case, making them challenging for time-constrained scenarios.

To address these issues, this paper proposes LEAP, an automated test method that uses \underline{LE}vy flight-based \underline{A}daptive \underline{P}article swarm optimization integrated with textual features to generate adversarial test cases. Specifically, we adopt Levy flight for population initialization to increase the diversity of generated test cases. We also design an inertial weight adaptive update operator to improve the efficiency of LEAP’s global optimization of high-dimensional text examples and a mutation operator based on the greedy strategy to reduce the search time.

We conducted a series of experiments to validate LEAP’s ability to test NLP software and found that the average success rate of LEAP in generating adversarial test cases is 79.1\%, which is 6.1\% higher than the next best approach (PSO$_{attack}$). While ensuring high success rates, LEAP significantly reduces time overhead by up to 147.6s compared to other heuristic-based methods. Additionally, the experimental results demonstrate that LEAP can generate more transferable test cases and significantly enhance the robustness of DNN-based systems.

\end{abstract}

\begin{IEEEkeywords}
NLP Software Testing, Particle Swarm
Optimization
\end{IEEEkeywords}

\section{Introduction}
In the field of NLP, Deep Neural Networks (DNNs) ($e.g.$, ELMo~\cite{peters-etal-2018-deep}, BERT~\cite{kenton2019bert}, GPT~\cite{brown2020language}, T5~\cite{raffel2020exploring}) have been developing rapidly in recent years. These networks are capable of extracting semantic, structural, and other information from text and have been widely integrated as new software components in safety-critical systems like market monitoring~\cite{liu2020improved}, code review~\cite{le2020deep}, and intelligence analysis~\cite{cho2020priority}. Such systems are referred to as \textit{DNN-based systems}. To address issues caused by malicious inputs, the software engineering (SE) community has proposed various techniques, including test coverage~\cite{lee2020effective, sparks2007automated, kolchin2022extending}, fuzz testing~\cite{guo2018dlfuzz, lemieux2018fairfuzz, xie2019deephunter}, and automated user interface testing~\cite{yu2019terminator, yousaf2019automated, li2019humanoid}. However, unlike software development methods that follow lifecycle frameworks~\cite{cusumano1995beyond, dima2018waterfall}, DNN-based systems do not require developers to design the system's rules. Instead, they rely on DNNs learning from large amounts of data to make decisions, which makes it challenging to ensure the robustness of DNN-based systems using traditional software testing methods.
Moreover, recent studies~\cite{szegedy2014intriguing, liang2018deep} have shown that DNN-based systems have significant robustness pitfalls due to the uninterpretability of such systems and the complexity of training data, as demonstrated by the following scenario.
\begin{figure}[t]
 \centering
\includegraphics[width=1.0\linewidth]{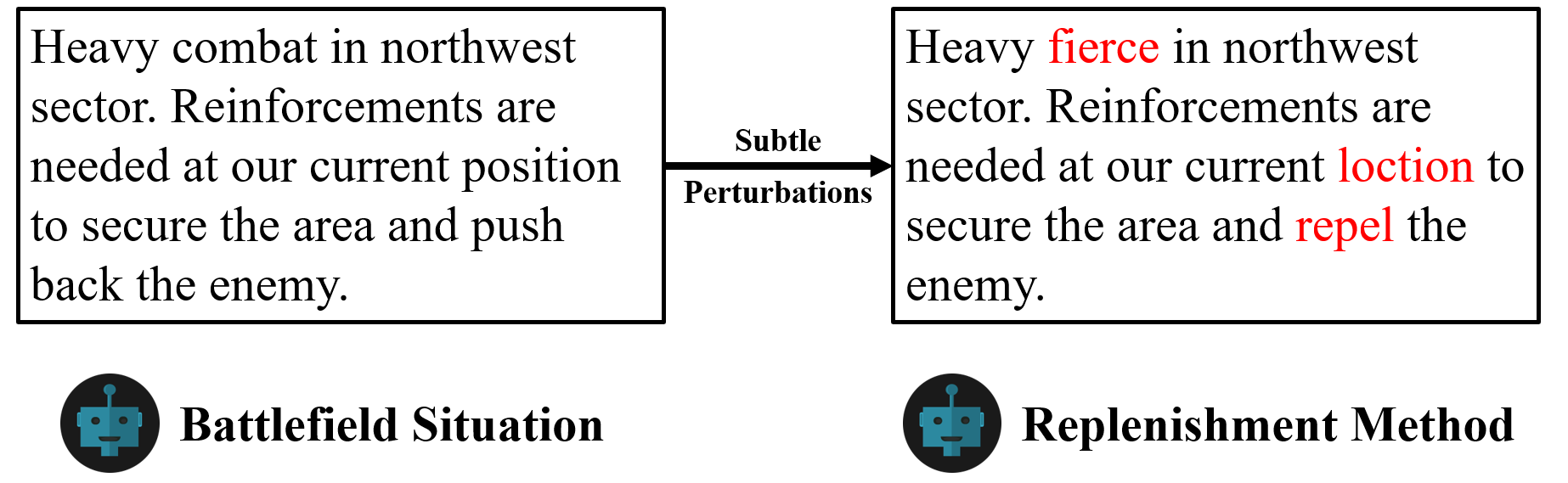}\\
 \caption{Subtle perturbed text (red) misleads military intelligence analysis systems to judge text labels from ``Battlefield Situation" to ``Replenishment Method".} 

 \label{Fig1}
 \vspace{-0.5cm}
\end{figure}

As shown in Fig.~\ref{Fig1},
the military intelligence analysis system is crucial to military information construction. It must classify a vast amount of text quickly to enhance intelligence analysis effectiveness and reduce command information loop cycles. However, when minor perturbations are added to the original intelligence, the system incorrectly classifies the text label as ``Replenishment Method" instead of ``Battlefield Situation." This error can result in valuable information being overlooked in the intelligence database, leading to missed fighting opportunities. Therefore, generating as many adversarial texts as possible as test cases is crucial to improving military intelligence analysis capabilities and advancing subsequent strategic deployments.
Since it is difficult to manually write numerous test cases for the DNN under test, which we refer to as the \textit{victim model}, in this paper, inspired by fuzz testing, we explore the potential of generating adversarial test cases~\cite{szegedy2014intriguing} in a heuristic manner to deceive DNNs' decision-making. This approach facilitates efficient detection of defects and vulnerabilities in NLP software.

We summarize the challenges faced by existing work as follows:

(1) \emph{Enhancing the ability to detect errors for DNN-based systems is the most urgent issue}. The testing process builds confidence in the system's quality by identifying and resolving defects. However, existing white-box and greedy strategy-based testing methods~\cite{2021Beyond,pruthi2019combating,yoo2021towards2} generate adversarial test cases based on a fixed perturbation paradigm, resulting in a low success rate of 0.4\% to 15.2\% on the commonly used AG's News dataset~\cite{zhang2015character} for toxic text detection tasks. Although heuristic testing methods~\cite{wang2019natural,zang-etal-2020-word} generate more successful test cases with a success rate of up to 70.5\% by iterating multiple times in an ample perturbation space, there is still room for improvement. Fig.~\ref{Fig1} illustrates the perturbation strategies of two existing works, including synonym replacement~\cite{wang2019natural} and character deletion~\cite{li2018textbugger}, which may generate syntactic errors when the replaced word has different part-of-speech tags or meanings. Such perturbations can be easily detected by syntactic-checking tools in software systems, leading to the generated test cases incapable of revealing errors in the system. A low success rate generates numerous invalid test cases, making testing methods difficult to work on small datasets.

(2) \emph{The existing methods take too much time to generate test cases.} Take the military software testing scenario in Fig. \ref{Fig1} as an example -- the rapid change of the battlefield situation requires the test methods to generate test cases quickly~\cite{hagen2013delivering}. Once the time limit of testing the victim model is exceeded, the generated test cases by the test method are useless for the improvement of the robustness of the victim model even if they can mislead the system's decision.
Although current heuristic testing methods~\cite{wang2019natural,zang-etal-2020-word} can generate more successful test cases,
the time of generating test cases for text sequences of length up to 250 is 58.53s (IMDB~\cite{maas2011learning}) and 177.81s (AG's News~\cite{zhang2015character}) on average, making them impractical for time and query-constrained scenarios.

To this end, we propose LEAP, an automated black-box testing method that employs PSO~\cite{kennedy1995particle} to search for adversarial test cases in NLP discriminative models. To increase the diversity of the population and improve the attack success rate of the test case, LEAP first generates the initial population using Levy flight and Brownian motion based on synonyms for each word, prepared using WordNet~\cite{10.1145/219717.219748}.
Next, as stated in the existing work~\cite{zhan2009adaptive} that the exponentially increasing perturbation space and complex search process require the search algorithm to have nonlinear search capability. Inspired by Shi et al.’s work~\cite{shi2001fuzzy}, we design a new adaptive inertia weight update strategy for LEAP to optimize the search path in an exponentially growing text space, which makes the search process more efficient. 
If LEAP fails to find any successful adversarial test case after each round of updating particles, a greedy mutation is performed to accelerate convergence.

In this paper, we investigate the ability of LEAP to generate adversarial test cases for three victim models on three datasets, including the classical LSTM model~\cite{liu2016recurrent2} and the two popular pre-trained models, BERT~\cite{kenton2019bert} and DistilBERT~\cite{2019DistilBERT}, with metrics including attack success rate~\cite{morris2020textattack2}, change rate~\cite{morris2020textattack2}, and perplexity score~\cite{jelinek1977perplexity}. 
We compared LEAP against different types of baselines, including gradient-based (i.e., A2T), greedy-based (i.e., Checklist and PRUTHI), and heuristic-based (i.e., PSO$_{attack}$ and IGA). Our results show that LEAP-generated test cases have the highest attack success rates with an average value of 79.1\% against 73.0\% for the next best approach (PSO$_{attack}$). Furthermore, LEAP consumes lower time overhead than other heuristic-based methods by 2.14s\textasciitilde147.57s. It thus can efficiently detect defects in the system. In addition, we conducted a transferability test, adversarial training, and an ablation study to further evaluate the performance of LEAP. We also assessed the naturalness of LEAP's test cases and found that it generates less modified and more natural test cases in most cases, as evidenced by the lower perplexity scores~\cite{jelinek1977perplexity}. 

The contributions of this paper include the following:
\begin{itemize}
\item We propose a new automated testing method, LEAP, which uses Levy flight~\cite{levy1939addition} along with Brownian motion to reasonably extend the perturbation range and improve the quality of adversarial test cases. During the iterative search in the perturbation space, LEAP utilizes the proposed adaptive algorithm and greedy mutation  for planning the search path to reduce the time overhead and query count. Our implementation and all raw data are open-source\footnote{https://github.com/lumos-xiao/LEAP}.

\item We conducted extensive experiments comparing LEAP with state-of-the-art automated testing methods for DNN-based NLP models. 
LEAP generated test cases with higher attack success rates while consuming less time.

\item We evaluated the effectiveness of adversarial test cases in improving the robustness of DNN-based systems. The experimental results show that adversarial training using LEAP's test cases can substantially (9.5\%\textasciitilde13.2\%) enhance the robustness of most victim models.
\end{itemize}

\section{Background}
\subsection{Problem Definition}
As a fundamental aspect of testing techniques for DNN-based NLP systems, 
the test data of a test case comprises a perturbed text sequence, and the expected result is the predicted label of the original text. LEAP performs automated testing on DNNs embedded in NLP software to generate adversarial examples as perturbed sequences of the test cases.
The notion of adversarial testing was introduced by Szegedy et al.~\cite{szegedy2014intriguing}. In this test method, a tester adds subtle perturbations $\varepsilon$ to the original data $x$, which can be digested by a machine learning model (i.e., the victim model) $f$, but is difficult for humans to perceive. This results in an adversarial example that can cause the victim model to produce erroneous results that differ from the original output $f(x)$. This paper focuses on generating test cases using black-box adversarial test methods, which only manipulate the inputs to the model.

LEAP uses the requirements of non-target adversarial testing as the objective function to find more test cases and test the DNNs more adequately. Specifically, given an original text segment $T_{ori}$ in the dataset and the corresponding adversarial test case $T_{adv}$, the optimization problem of LEAP can be defined as
\begin{equation}
\begin{aligned}
& \underset{T_{a d v} \in C\left(T_{o r i}\right)}{\arg \min }\left\|T_{o r i}, T_{a d v}\right\| \\
& \text { s.t. } F\left(T_{o r i}\right) \neq F\left(T_{a d v}\right)
\end{aligned}
\end{equation}
where $\|a, b\|$ denotes the difference between two pieces of text segments $a$ and $b$, such as change rate, embedding distance, etc; $F$ denotes the victim model; $C$ denotes LEAP's constraint on the quality of the adversarial test cases, here including the stop word filter~\cite{li2020bert} and the maximum change rate limit, because an excessive change rate affects the semantics and naturalness of the generated cases.

\subsection{Particle Swarm Optimization}
PSO is a population collaborative-based search algorithm developed by Kennedy and Eberhart~\cite{kennedy1995particle} in 1995. It simulates the foraging behavior of a flock of birds, where each individual is called a particle. It has been successfully applied in many fields, such as economic management~\cite{phommixay2020review}, information science~\cite{latchoumi2019bio}, engineering technology~\cite{su2022sewage} and emotional binary classification in NLP~\cite{tambouratzis2019pso}.
In the original PSO, the particles simulate the solution of the optimization problem in the search space. The fitness value of a particle is evaluated according to its position, usually  in terms of the objective function or optimization problem, and the particle velocity is a vector indicating the direction and distance it will move. The PSO process is described as follows:

(1) Initialization. A random population of particles is generated, and the initialization involves randomly generating each particle's position and velocity vector.

(2) Evolutionary iteration. Each particle searches the entire solution space by updating its velocity and position according to its optimal position $lBest$ so far and the optimal position $gBest$ of the population. When the particle population position is updated, the particle's optimal position and the population's optimal position are also updated.

(3) Iteration termination. When the iteration termination condition is met, the algorithm stops searching, and the last optimal position searched is the optimal solution.

In the evolutionary iteration, the updated equation for the velocity $v_d^n$ of the $n$-th particle in $d$ directions is
\begin{equation}
v_d^n=w v_d^n+c_1 * r_1 *\left(lBest_d^n-x_d^n\right)+c_2 * r_2 *\left(gBest_d^n-x_d^n\right)
\end{equation}

The position update equation of the particle is 
\begin{equation}
x_d^n=x_d^n+v_d^n
\end{equation}
where $x_d^n$ denotes the $d$-th dimension of the $n$-th particle in the current population; $w$ is the inertia weight; $c_1$ and $c_2$ are learning factors; $r_1$ and $r_2$ are random numbers uniformly distributed in the range of [0,1].

The setting of control parameters tremendously influences the performance of PSO~\cite{shi2001fuzzy}. The parameters $c_1$ and $r_1$ indicate the degree of influence of particles by $lBest$, i.e., how the particles assess their own information sharing and cooperation with other particles in the current population; $c_2$ and $r_2$ indicate the degree of influence of 
other particles by $gBest$, i.e., how the particles assess the information sharing and cooperation of other particles. The inertia weight determines the succession to the current velocity of the particle~\cite{li2021adaptive}. 

For the iteration termination, there are two general termination conditions: (1) the current iteration number $t$ reaches the preset maximum iteration number; or (2) there are individuals in the population that satisfy the accuracy requirements of the optimization problem. 

\section{Design of LEAP}

\begin{figure}[t]
 \centering
\includegraphics[width=0.85\linewidth]{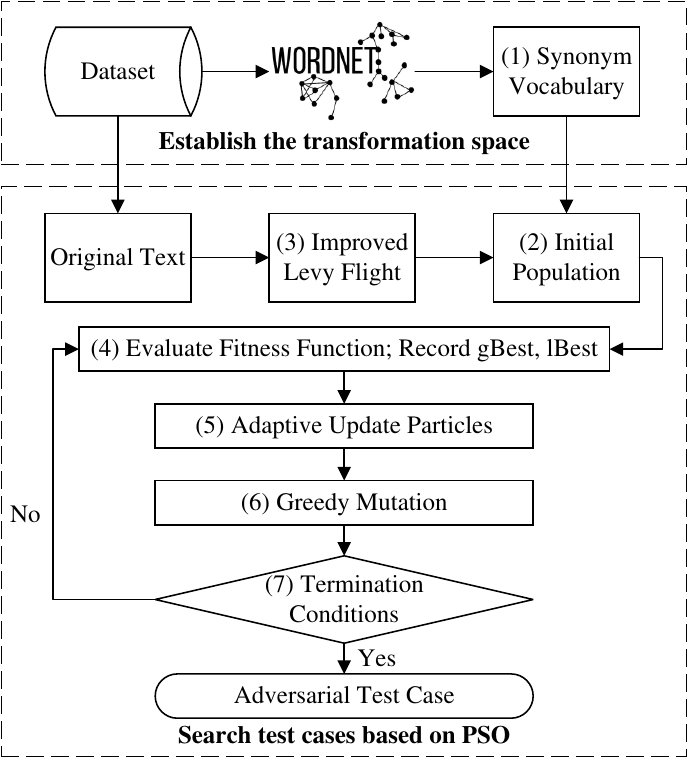}\\
 \caption{Overview of LEAP.}\label{Fig2}
 \vspace{-0.5cm}
\end{figure}

Fig.~\ref{Fig2} overviews the proposed LEAP, which aims to generate adversarial test cases using actual examples from the test dataset. (1) It begins by counting all the words in the dataset and using a synonym lexicon called WordNet~\cite{10.1145/219717.219748} to find synonyms for each word. (2) It then selects an original text sequence from the dataset and replaces a word with its synonym to obtain the initial position. The initial velocity (3) is obtained through a modified Levy flight. The initial position and velocity together determine the initial population of particles. Next, LEAP (4) performs an iterative search, using the confidence score of the victim model as the fitness function.
It then (5) adaptively updates the velocity and position of the particles. LEAP also (6) performs greedy mutation based on the change rate and fitness score. Suppose the best individual in the population (7) satisfies the termination conditions, which include successfully changing the prediction of the original text and reaching the maximum number of iterations. In such a case, the output is an adversarial test case. Otherwise, the iteration continues. 

 The depiction of LEAP is divided into two parts: 1) establishing the transformation space and 2) searching test cases based on PSO. 
 
\subsection{Establishing the transformation space}
To heuristically search for adversarial test cases, LEAP first defines the search space. Given that the original text $T_{ori}$=\{$w_1$, $w_2$, ..., $w_n$\} contains $n$ words, LEAP generates potential test cases $T_{ori}'$ by replacing a word $w_i$ in $T_{ori}$ with its synonym $w_i'$, and multiple $T_{ori}'$s for each original text $T_{ori}$ constitute the search space of the test dataset together. LEAP focuses on generating semantically correct test cases and therefore uses WordNet to construct a synonym vocabulary for each word in the dataset. WordNet is a broad-coverage English lexical-semantic network where nouns, verbs, adjectives, and adverbs are respectively organized into a network of related words, with each set of synonyms representing a basic semantic concept and various relations connecting these sets.


The process of generating a synonym vocabulary in LEAP using WordNet is superior to other methods, such as using word embedding~\cite{wang2019natural}, language model~\cite{li2020bert}, and HowNet~\cite{zang-etal-2020-word}, this is because:
\begin{itemize}
\item The word embedding method can find many candidate words by changing the embedding distance threshold to ensure diversity in the search space. However, it also introduces low-quality substitutions, such as lexical errors.
\item The method using language models to build the search space produces fluent sentences because these models (especially pre-trained models~\cite{kenton2019bert,brown2020language}) are obtained from large text datasets and contain contextual semantic knowledge. However, they are prone to syntactic errors because linguistic features such as syntax and semantics are ignored.
\item HowNet~\cite{1276017} is an extensive dictionary that uses ``sememe" to describe words and semantics. Different from WordNet, it only considers synonymy and positive and negative colors in semantic relations, ignoring the summary of related words, such as antonyms of words. The search space established using HowNet is too small, reducing the population diversity of PSO and thereby affecting the algorithm's ability to find higher-quality test cases.
\end{itemize}

We thus use WordNet to generate a synonym vocabulary for LEAP. The output of WordNet is a list of candidate words that are the synonym of each word $w_{i}$ in the original text $T_{ori}$.

\subsection{Searching test cases based on PSO}
The respective synonym vocabulary for each word in the original text forms the search space of LEAP, which approaches automated testing as a combinatorial optimization problem and uses our improved PSO to find adversarial test cases that satisfy the objective function and constraints within the search space. We improve PSO since the original one is only suitable for continuous search spaces, but the perturbation space for the NLP test case generation task is discrete, 
LEAP thus updates PSO by probability according to the scalar shift discussed in Section \ref{sec3-2-2} inspired by~\cite{zang-etal-2020-word}.
In addition, it improves PSO using Levy flight and adaptive methods to generate higher-quality adversarial test cases with less time overhead. Algorithm \ref{algorithm1} outlines the search process. Next, we detail this algorithm.
\begin{algorithm}[t]
       \footnotesize
	\caption{Search Process in LEAP}
	\begin{algorithmic}[1]
		\REQUIRE $T_{ori}$: Original text, $max\_iters$: Max iteration, $pop\_size$: Number of the population in each iteration.
		\ENSURE $T_{adv}$: Adversarial test case.
        \STATE $T_{pop}$$\leftarrow$Levy-Initialization$(T_{ori})$ via $Eq.7$;
        \IF{$T_{adv}$ in $T_{pop}$}
        \RETURN $T_{adv}$
        \ENDIF
        \STATE $gBest$=max\{$T_{pop}$\};
        \STATE $lBest$=copy\{$T_{pop}$\};
        \WHILE{not exceed $max\_iters$}
        \STATE Adaptively set inertia weight $\omega$ via $Eq.8$;
        \FOR{$n$ in $pop\_size$}
        \STATE Update the velocity and position of particle $n$;
        \ENDFOR
        \STATE Evaluate current population;
        \STATE Greedy-Mutation based on change rate via $Eq.11$;
        \FOR{$n$ in $pop\_size$}
        \IF{fit($n$)$>$fit($lBest$)}
        \STATE $lBest$=$pop_{n}$;
        \ENDIF
        \ENDFOR
        \IF{fit($lBest$)$>$fit($gBest$)}
        \STATE $gBest$=$lBest$;
        \ENDIF
        \STATE Evaluate current population;
        \ENDWHILE
        \RETURN $T_{adv}$$\leftarrow$$gBest$
	\end{algorithmic}\label{algorithm1}
\end{algorithm}

\subsubsection{Population initialization based on Levy flight}
\
\newline
The main task of population initialization is to determine the initial velocity and position of the particles to perform the search. 
To achieve this, LEAP uses the confidence of the victim model as the fitness of the particles, since the aim of an adversarial test method is to create inputs that can fool the model into making incorrect predictions with high confidence. By using confidence as the fitness function, the optimization algorithm can search for inputs that are most likely to be misclassified.
LEAP generates the initial position based on the fitness score. Specifically, for each word $w_{i}$ in the original input text sequence $T_{ori}$, 
LEAP replaces only one word with its synonym at a time to minimize the modification of $T_{ori}$. This generates a series of new sequences to construct the 
search space corresponding to the current input. LEAP then traverses the search space to find the example with the highest fitness score, which flips the original prediction.
The replaced synonym word in LEAP is designated as the best neighbor of the original word. 
For an original sequence $T_{ori}$, the new sequences generated from $T_{ori}$ by replacing different words with their synonym have different fitness values.
LEAP thus uses these fitness values as probabilities associated with each new sequence. Based on the probabilities, it randomly selects a word in $T_{ori}$ and uses the best neighbor of this word to replace it. The replaced text is the initial position of the particle.

\begin{figure}[t]
 \centering
\includegraphics[width=1.0\linewidth]{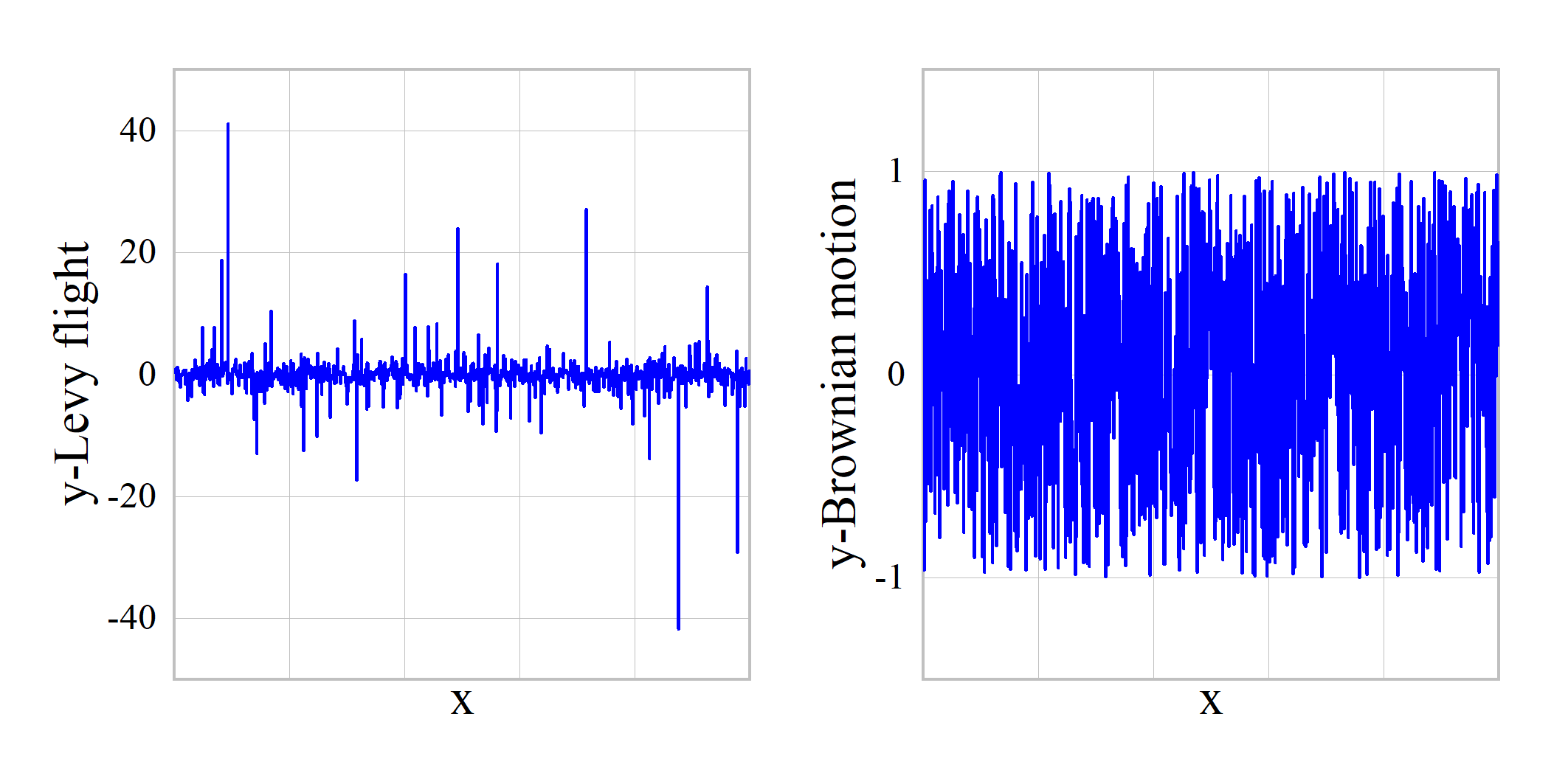}
 \caption{Comparison of Levy flight and Brownian motion. x represents the number of steps performed and y represents the step length. Obviously, the search area covered by Levy flight (in range [-40, 40]) is much broader than Brownian motion (in range [-1, 1]). }\label{Fig3}
  \vspace{-0.5cm}
\end{figure}

In ~\cite{zang-etal-2020-word} that uses PSO to search for adversarial test cases, the velocity of the particles is initialized using Brownian motion~\cite{einstein1956investigations}, which focuses on local search. However, the search space for the NLP test case generation task increases exponentially as the number of words in the input case increases. This search process
is prone to get stuck in local optima. To address this issue, LEAP uses Levy flight to initialize the velocity of the particles (Lines 1-6). Levy flight is a random wandering mode proposed by French mathematician Paul Pierre Levy in 1930s~\cite{levy1939addition}, in which the steps follow the Levy distribution and can move in multidimensional space with isotropic random directions. Fig. \ref{Fig3} illustrates the difference between Levy flight and Brownian motion. Within 500 steps, the step length of Brownian motion mainly bounces around the current point in a small area, 
while Levy flight has a wandering characteristic that combines short walks and long jumps.
This means that Levy flight has a higher probability of taking long steps than normal random walks. 
In the context of NLP, this can be useful for exploring a larger potential search space, which can improve the chances of finding effective adversarial test cases. Specifically, Levy flight allows the population to explore a wider range of input space, leading to more diverse populations. A diverse population increases the chances of finding effective adversarial test cases and helps to avoid local optima.

The step size of the Levy flight is determined by the Levy distribution, which is complex and has not been implemented yet. It is thus usually simulated using the Mantegna algorithm~\cite{mantegna1994fast} with a step size $s$ calculated by:
\begin{equation}
s=\frac{\mu}{|v|^{1 / \beta}}\label{eq3}
\end{equation}
where $\mu \sim N\left(0, \sigma_\mu^2\right)$,$ \quad v \sim N\left(0, \sigma_v^2\right)$, $\beta$ usually takes the value 1.5, and
\begin{equation}
\sigma_\mu=\left\{\frac{\Gamma(1+\beta) \sin \left(\frac{\pi \beta}{2}\right)}{\Gamma\left[\frac{(1+\beta)}{2}\right] \beta^2 \frac{(\beta-1)}{2}}\right\}^{1 / \beta}
\end{equation}
\begin{equation}
\sigma_v=1
\label{eq5}
\end{equation}

LEAP randomly generates the Brownian motion's step size, and each particle's initial velocity is obtained by combining Levy flight and Brownian motion. The assignment formula is:
\begin{equation}
v_{\text {init }}=\left\{\begin{array}{l}
\operatorname{levy}\left(\beta, \sigma_v\right) \quad, \operatorname{levy}\left(\beta, \sigma_v\right)>\operatorname{rand}\left(v_{\min }, v_{\max }\right) \\
\operatorname{rand}\left(v_{\min }, v_{\max }\right), \text { others }
\end{array}\right.
\end{equation}
It is observed that the step size of Brownian motion is broader than that of Levy flight, which almost occurs when both values are small. The minor oscillation feature of Brownian motion makes it have better local search capability, so the value generated by Brownian motion is used in this case. The rest of the cases use the step size generated by the Levy flight to enhance the global search capability of LEAP and thus generate better-quality adversarial test cases.
\subsubsection{Adaptive update particles}\label{sec3-2-2}
\ 
\newline
If there are no test cases in the initial population of LEAP that can test successfully, the population will be iterated, with the velocity of the particles being adaptively updated first, and then the particles being shifted according to the velocity (Lines 8-11). Balancing global and local search by adjusting the step size is vital for the success and efficiency of the iterative search in heuristic algorithms. PSO uses inertia weights to balance global and local search capabilities, with larger weights contributing to global search and smaller weights contributing to local search. Changing the inertia weights allows for dynamic adjustment of the search capability. The existing method~\cite{zang-etal-2020-word} uses a linearly decreasing inertia weight to dynamically adjust the search process so that PSO has more global search capability at the beginning and more local search capability near the end of the run. 
However, the search space increases exponentially with the number of replaced words, which means that the search process of LEAP is non-linear and requires tremendous time overhead. Besides, the method of linearly decreasing inertia weights has a linear transition of search capability from global to local search, resulting in it easily falling into the saddle of high-dimensional text space later in the search.
Therefore, the inertia weights should be nonlinear and change dynamically to provide a better dynamic balance between global and local search capabilities and achieve better performance.
\begin{equation}
\omega_n^i= \begin{cases}\omega_{\min }+\frac{\left(f i t_n^i-f i t_{\min }^i\right)\left(f i t_{\max }^i-f i t_{\min }^i\right)}{f i t_{\text {mean }}^i-f i t_{\min }^i}, & f i t_n^i<f i t_{\text {mean }}^i \\ \operatorname{levy}\left(\beta, \sigma_v\right) \in\left(\omega_{\text {mean }}, \omega_{\max }\right), &  \text { others }\end{cases}\label{eq7}
\end{equation}

LEAP uses a new adaptive inertia weight update method, as shown in Equation \ref{eq7}, where $\omega_{\min }$ and $\omega_{\max }$ are hyperparameters. Suppose the fitness score of the $n$-th particle in the $i$-th generation is less than the average value of all fitness scores. In that case, this particle can be considered far from the actual value or stuck in a local search. Then, its inertia weight is adaptively adjusted based on the fitness score.
Otherwise, the inertia weight is the value generated by Levy flight, ensuring that the search process has certain randomness and explores a larger perturbation space. After obtaining the new inertia weights, the velocity is updated according to Equation \ref{eq8}.
\begin{equation}
v_d^n=\omega^n v_d^n+v_{\max }\left(1-\omega^n\right)\left[I\left(lBest^i, x_n^i\right)+I\left(gBest, x_n^i\right)\right]\label{eq8}
\end{equation}
In order to search the discrete perturbation space, where
\begin{equation}
I(a, b)=\left\{\begin{array}{l}
1, a=b \\
-1, a \neq b
\end{array}\right.
\end{equation}

The update of the position is similarly divided into two steps. In the first step, a new move probability $P_1$ is introduced by which a particle determines whether to move to its individual best position; in the second step, each particle determines whether to move to the global best position with another move probability $P_2$. The change of each position dimension depends on $softmax(v_d^n)$. $P_1$ and $P_2$ are hyperparameters that change with iteration to improve the search efficiency by adjusting the balance between local and global search.

\subsubsection{Greedy mutation}
\ 
\newline
In biology, genetic mutations result in differences among individuals within a population, in terms of their structure and function. To simulate this process and ensure population diversity, LEAP introduces a mutation operator to the original PSO algorithm (Line 13). To prevent excessive modification of the text, LEAP generates variation probabilities based on the change rate (\emph{C-rate}) of the current particle from the original text, as shown in Equation \ref{eq10}. 
\begin{equation}
p_{\text {mutation }}=1-\gamma \cdot \emph{C-rate}\label{eq10}
\end{equation}

Randomness is ensured by comparing the mutation probability with a random number in the range [0,1). If the generated random value is less than $p_{\text {mutation}}$, greedy mutation is performed on the particle: the words in the text sequence are replaced one by one to find the perturbed position that makes the greatest improvement in the fitness score, and then the original particle is replaced using the perturbed text. Next, LEAP updates $gBest$ and $lBest$ by fitness score, and $gBest$ is output as an adversarial test case when the iteration terminates.


\section{Experiment Setup}
We have conducted a series of experiments on three text classification datasets and three deep learning models to validate the performance of LEAP in generating test cases. We have made LEAP and all raw data publicly available. All experiments were conducted on an Ubuntu 18.04.5 LTS server with NVIDIA RTX A4000, a 12-core 2.20GHz processor Intel(R) Xeon(R) Gold 5320, and 32GB physical memory. We conducted three repetitions of experiments and averaged the experiment results for each metric. Similar to many well-acknowledged studies~\cite{zang-etal-2020-word,jin2020bert, wang2019natural,alzantot2018generating2}, the victim models were tested on a set of 1,000 randomly selected examples in each experiment. Therefore, it is believed that this experimental scale is sufficient to cover different input data types and ensure the representativeness and credibility of the experiment results.

\subsection{Hyperparameters}
LEAP is a heuristic testing method based on PSO, and the selection of hyperparameters significantly influences its performance. Among them, the population size ($pop\_size$) determines the coverage of the discrete text space, and the maximum number of iterations ($max\_iters$) affects the computational cost required for the search process. The inertia weight ($\omega$) and acceleration coefficients ($P_1$, $P_2$) jointly determine the breadth and depth of the search; overly large or small values of these hyperparameters may cause LEAP to get trapped in local optima. By parameter tuning, we set the number of individuals in the particle swarm to 60, the maximum number of iterations to 20, and the hyperparameters $\omega_{\min }$, $\omega_{\max }$, $P_1$, $P_2$ and $\gamma$ 0.2, 0.8, 0.8, 0.2 and 1, respectively.





\subsection{Datasets}
\textit{IMDB}\footnote{https://s3.amazonaws.com/fast-ai-nlp/imdb.tgz}~\cite{maas2011learning}. A dataset for emotional binary classification containing 50,000 positive and negative movie reviews was grabbed from online sources. The average length of each sequence is 215.63 words. It is divided into two parts, namely 25,000 training reviews and 25,000 test reviews. Their polarization characterizes these movie reviews.

\textit{AG's News}\footnote{https://s3.amazonaws.com/fast-ai-nlp/ag\_news\_csv.tgz}~\cite{zhang2015character}. This dataset quotes 496,835 news articles from more than 2,000 news sources in the 4 categories of AG's News Corpus (World, Sports, Business, and Science/Technology) in the title and description fields. We concatenate the title and description fields of each news article and use the dataset organized by kaggle\footnote{https://www.kaggle.com/amananandrai/ag-news-classification-dataset}, in which each category contains 30,000 training examples and 1,900 test examples. Each example contains an average of 43 words.

\textit{Poem Sentiment(POEM)}\footnote{https://github.com/google-research-datasets/poem-sentiment}~\cite{sheng2020investigating}. This dataset contains 3,085,117 lines of poetry from hundreds of Project Gutenberg books, which can be used for tasks such as sentiment classification. Each line has a corresponding Gutenberg ID (1,191 unique values) from Project Gutenberg. These text segments are divided into four categories, with an average length of 8 words per segment.

\subsection{Victim models}
To evaluate the test performance of LEAP on different DNN-based systems, we choose BERT~\cite{kenton2019bert} and its concise scheme Distil-BERT~\cite{2019DistilBERT}, thus verifying the performance of researchers' most common NLP models. We also report experimental results on a 
LSTM for text classification~\cite{liu2016recurrent2}, which is widely used as a classical deep learning model with excellent performance before the advent of pre-trained models. By parameter tuning, the number of hidden layer neurons of TextBiRNN was set to 150; the dropout ratio was set to 0.1, and the maximum sequence length was set to 250. All these models have been pre-trained on BookCorpus~\cite{zhu2015aligning}, a dataset consisting of 11,038 unpublished books and English Wikipedia (excluding lists, tables, and titles). We also finetuned the bert-base-uncased\footnote{https://huggingface.co/bert-base-uncased}, distilbert-base-uncased\footnote{https://https://huggingface.co/distilbert-base-uncased} models published by Hugging Face for each dataset.

\subsection{Baselines}
We investigated the recent works in terms of the testing framework~\cite{morris2020textattack2,tan2021reliability}, degree of automation~\cite{ribeiro2022adaptive,wu2019errudite}, and application scenario~\cite{li2018textbugger,li2020bert,jin2020bert,ye2022texthoaxer}. Among these, we selected the testing framework Textattack~\cite{morris2020textattack2}, which does not require manual intervention, and conducted experiments in the context of soft-label black-box testing. To compare LEAP with different fully automated testing methods, we implemented four popular black-box testing methods and one state-of-the-art white-box testing method. Specifically, these methods are:

1) \emph{IGA} proposed by Wang et al.~\cite{wang2019natural}: the fitness function consists of confidence and alienation rate. Using single-point crossover, the text of the two parents is randomly cut to merge into a new text segment. Allowing to replace the words that have been replaced before, to a certain extent, avoids falling into the trap of local optima.

2) \emph{PSO$_{attack}$} proposed by Zang et al.~\cite{zang-etal-2020-word}: a word-level automated testing method which reforms in two steps -- reducing search space and searching for adversarial test cases through designing a word substitution method based on sememes, 
and presenting a search algorithm based on particle swarm optimization.

3) \emph{CheckList} proposed by Ribeiro et al.~\cite{2021Beyond}: inspired by principles of behavioral testing in software engineering, CheckList guides users in what to test by providing a list of linguistic capabilities. To break down potential capability failures into specific behaviors, CheckList introduces different test types and then implements multiple abstractions 
to generate adversarial test cases.

4) \emph{PRUTHI} proposed by Pruthi et al.~\cite{pruthi2019combating}: explores adversaries which perturb sentences with four types of character-level edits: (1) Swap: swapping two adjacent internal characters of a word. (2) Drop: removing an internal character of a word. (3) Keyboard: substituting an internal character with adjacent characters of QWERTY keyboard (4) Add: inserting a new character internally in a word. 

5) \emph{A2T} proposed by Yoo et al.~\cite{yoo2021towards2}: the component of this method is designed to generate adversarial test cases with lower computational cost, which is accelerated by making two key choices when constructing the test: (1) DistilBERT semantic textual similarity constraint, and (2) a cheaper gradient-based word importance ranking white-box method.
\subsection{Evaluation measures}
We choose five evaluation indicators for the experiment:

1) \emph{Success rate} (\emph{S-rate})~\cite{morris2020textattack2} of generated adversarial test cases among all targeted text segments. In this experiment, its formula can be expressed as follows:
\begin{equation}
\text{S-rate}=\frac{N_{a d v}}{N}
\end{equation}
where, $N_{a d v}$ is the number of adversarial test cases that test victim models successfully, and $N$ is the total number of input examples ($N$ = 1,000 in our experiment) for the current test method.

2) \emph{Change rate} (\emph{C-rate})~\cite{morris2020textattack2}, which represents the average proportion of the changed words in the original text. C-rate can be expressed as:
\begin{equation}
\text {C-rate }=\frac{1}{N_{a d v}} \sum_{k=1}^{N_{adv}} \frac{\operatorname{diff} T_k}{\operatorname{len}\left(T_k\right)}
\end{equation}
where $\operatorname{diff} T_k$ represents the number of words replaced in the input text $T_k$ and $\operatorname{len}$($*$) represents the sequence length. C-rate is an indicator designed to measure the difference in content between the generated test cases and the original examples.

3) \emph{Perplexity} (\emph{PPL})~\cite{jelinek1977perplexity}, an indicator used to assess the fluency of textual test cases. Perplexity is defined as the exponentiated average negative log-likelihood of a sequence. If we have a tokenized sequence $X$=($x_0$,$x_1$,\dots,$x_t$), then the perplexity of $X$ is,
\begin{equation}
\operatorname{PPL}(X)=\exp \left\{-\frac{1}{t} \sum_i^t \log p_\theta\left(x_i \mid x_{<i}\right)\right\}
\end{equation}
where $\log p_\theta\left(x_i \mid x_{<i}\right)$ is the log-likelihood of the $i$-th token conditioned on the preceding tokens $x_{<i}$ according to the language model~\cite{meister2021language}. Intuitively, given the language model for computing PPL, the more fluent the test case, the less confusing it is.

4) \emph{Time overhead} (\emph{T-O})~\cite{morris2020textattack2}, which refers to the average time it takes for a test method to generate a successful test case.

5) \emph{Query number} (\emph{Q-N})~\cite{morris2020textattack2}, which indicates the average number of times a population-based method needed to query the victim model when generating a test case. The query number and the time overhead together reflect the efficiency of the testing method.

We use C-rate and PPL to quantitatively measure the naturalness and similarity  between adversarial test cases and original ones, as both are easier to reproduce than human evaluation.
Regarding time overhead and query number, we compare LEAP with IGA and PSO$_{attack}$, which are also heuristic testing methods, considering that non-heuristic test methods~\cite{2021Beyond,pruthi2019combating,yoo2021towards2} generate test cases much faster due to the different search strategies.
However, the experimental results in Section \ref{sec5-1}
show that the quality of test cases generated by such methods is much inferior to that of heuristic methods.
\subsection{Definition of robustness}
IEEE~\cite{iso2017iso} defines the robustness in software engineering as \textquotedblleft degree to which a system, product or component performs specified functions under specified conditions for a specified period of time\textquotedblright. Similar to~\cite{wang2022measure}, we define robustness as follows: denoting the input as $x$ and the relevant gold label for the main task as $y$, assuming that a model $f$ is trained on $(x, y) \sim \mathcal{D}$. Now given the adversarial test case $\left(x^{\prime}, y^{\prime}\right) \sim \mathcal{D}^{\prime}\neq\mathcal{D}$, we can measure the robustness of the model by the prediction results of $f$ on $\left(x^{\prime}, y^{\prime}\right)$. 
Compared to the raw prediction accuracy on $\mathcal{D}$, the less the model's prediction accuracy on $\mathcal{D}^{\prime}$ drops, 
the fewer test cases the model misclassifies, the more robust it is.  

\begin{table*}[th]
\centering
\caption{Performance of six methods to generate test cases. $nan$ in CheckList refers to the predictions of all generated test cases being the same as the original labels.
}
\label{tab1}
\begin{tabular}{c|c|ccc|ccc|ccc}
\hline
\multirow{2}{*}{\textbf{Dataset}} & \multirow{2}{*}{\textbf{Baseline}} & \multicolumn{3}{c|}{\textbf{BERT}}                   & \multicolumn{3}{c|}{\textbf{Distil-BERT}}            & \multicolumn{3}{c}{\textbf{LSTM}}               \\ \cline{3-11} 
                                  &                                    & \textit{S-rate} & \textit{C-rate} & \textit{PPL}     & \textit{S-rate} & \textit{C-rate} & \textit{PPL}     & \textit{S-rate} & \textit{C-rate} & \textit{PPL}     \\ \hline
\multirow{6}{*}{\textbf{IMDB}}    & \textbf{PSO$_{attack}$}                 & 0.913           & 0.174           & 82.836           & 0.902           & 0.166           & 239.405          & 0.832           & 0.025           & 42.003           \\
                                  & \textbf{IGA}                       & 0.908           & 0.123           & 49.584           & 0.892           & 0.121           & 58.197           & 0.799           & 0.121           & 44.165           \\
                                  & \textbf{Checklist}                 & 0.020           & 0.224           & 46.291           & 0.016           & 0.611           & 46.012           & 0.188           & 0.407           & 64.049           \\
                                  & \textbf{A2T}                       & 0.246           & 0.083           & 270.717          & 0.304           & 0.068           & 45.541           & 0.668           & 0.043           & 39.628           \\
                                  & \textbf{PRUTHI}                    & 0.134           & \textbf{0.046}  & 207.779          & 0.188           & \textbf{0.034}  & 43.227           & 0.224           & \textbf{0.005}  & 45.263           \\
                                  & \textbf{LEAP}                      & \textbf{0.922}  & 0.113           & \textbf{43.603}  & \textbf{0.910}  & 0.115           & \textbf{42.179}  & \textbf{0.860}  & 0.040           & \textbf{36.364}  \\ \hline
\multirow{6}{*}{\textbf{AG's News}}    & \textbf{PSO$_{attack}$}                 & 0.696           & 0.244           & 690.094          & 0.644           & 0.248           & 800.665          & 0.692           & 0.197           & 343.657          \\
                                  & \textbf{IGA}                       & 0.705           & 0.179           & 1013.61          & 0.621           & 0.168           & 750.969          & 0.656           & 0.165           & 305.628          \\
                                  & \textbf{Checklist}                 & 0.004           & 0.032           & 1555.637         & 0.008           & \textbf{0.029}  & 1677.115         & 0.036           & 0.066           & 445.628          \\
                                  & \textbf{A2T}                       & 0.096           & 0.091           & 1142.844         & 0.076           & 0.090           & 925.113          & 0.152           & 0.077           & 512.617          \\
                                  & \textbf{PRUTHI}                    & 0.092           & \textbf{0.029}  & 1202.431         & 0.064           & 0.031           & 1182.057         & 0.104           & \textbf{0.031}  & 423.197          \\
                                  & \textbf{LEAP}                      & \textbf{0.812}  & 0.157           & \textbf{673.893} & \textbf{0.672}  & 0.162           & \textbf{744.605} & \textbf{0.896}  & 0.211           & \textbf{214.954} \\ \hline
\multirow{6}{*}{\textbf{POEM}}    & \textbf{PSO$_{attack}$}                 & 0.658           & 0.196           & 2176.741         & 0.640           & 0.201           & \textbf{620.848} & 0.595           & 0.165           & 616.108          \\
                                  & \textbf{IGA}                       & 0.576           & 0.179           & 2198.094         & 0.588           & 0.200           & 3921.968         & 0.499           & 0.143           & 523.979          \\
                                  & \textbf{Checklist}                 & \textit{nan}    & \textit{nan}    & \textit{nan}     & \textit{nan}    & \textit{nan}    & \textit{nan}     & 0.048           & 0.051           & 503.804          \\
                                  & \textbf{A2T}                       & 0.100           & 0.169           & \textbf{718.661} & 0.165           & 0.167           & 729.203          & 0.141           & 0.063           & 552.704          \\
                                  & \textbf{PRUTHI}                    & 0.469           & 0.168           & 2472.985         & 0.416           & 0.162           & 5795.784         & 0.129           & \textbf{0.025}  & \textbf{448.347} \\
                                  & \textbf{LEAP}                      & \textbf{0.714}  & \textbf{0.161}  & 2076.027         & \textbf{0.681}  & \textbf{0.157}  & 991.764          & \textbf{0.651}  & 0.133           & 489.931          \\ \hline
\end{tabular}
 \vspace{-0.5cm}
\end{table*}

\section{Experiment Results and Analysis}
In this section, we present five research questions and discuss the experimental results.

\vspace*{0.2em}\noindent\textbf{RQ1: How is the quality of the generated test cases by LEAP for different victim models and datasets?}\label{sec5-1}\vspace*{0.2em}

To evaluate LEAP, we compare its success rate, change rate, and perplexity with other baselines.
Table \ref{tab1} shows the comparison results on different datasets and victim models.

Compared to all the baselines, LEAP achieves higher success rates for each dataset and victim model, especially on the multi-categorical and long-series dataset, i.e., AG's News. When generating test cases for BERT, LEAP achieves a success rate of 81.2\% compared to the baseline success rates of 69.6\%, 70.0\%, 0.4\%, 9.6\%, and 9.2\%, which implies that LEAP can test more thoroughly against DNN-based systems with robust performance. In terms of change rate, LEAP achieves optimal results in only a few cases, with PRUTHI and CheckList often having better performance because these two methods have strict restrictions on the modification of the original text and therefore sacrifice too much performance in success rate. In the experiments tested on Distil-BERT finetuned by IMDB, the change rates of Checklist, PRUTHI, and LEAP are 61.16\%, 3.4\%, and 11.5\%. However, the success rate of the three is 1.6\%, 18.8\%, and 91\%, respectively, the  disparity of which is significant. In addition, LEAP's PPL scores are the lowest for most cases, indicating that LEAP can generate more fluent and natural test cases. Even though LEAP's PPL is not the lowest in a few cases, it guarantees a sufficiently high success rate. For example, when testing a bidirectional LSTM trained by Poem Sentiment, LEAP outperforms PRUTHI by 52.2\% in terms of success rate, while PRUTHI achieves a slightly better PPL score than LEAP.

In addition, Table \ref{tab1} shows that the three heuristics of PSO$_{attack}$, IGA, and LEAP always have the highest success rate. Besides, Table \ref{tab2} presents test cases generated from the same testing sequence by the three methods on a BERT finetuned by AG's News. It can be seen that the test case generated by LEAP not only deceives the victim model with high confidence but also makes minor and more natural changes to the original text. On the other hand, although LEAP and PSO$_{attack}$ also chose PSO for the iterative search, the adversarial test case generated by LEAP shows better text quality regarding the change rate and PPL score.
\begin{table}[!htbp]
\centering
\caption{Examples of adversarial test cases generated by three methods using BERT as the victim model.
}\label{tab2}
\setlength{\abovecaptionskip}{0.cm}
\setlength{\belowcaptionskip}{-0.cm}
\centering
\renewcommand\arraystretch{1.5}
\footnotesize{
\begin{tabular}{l}
\hline
(Original Text)   Prediction = \textbf{Sci/Tech}. (Confidence = \textbf{0.983})                                                                                                                                                                                                                                                                                        \\ \hline
TheStreet.com May Be Up for Sale --   Report (Reuters) Reuters - The\\Street.com Inc. , the financial news and   commentary Web site, may be\\up for sale, according to a report in Business Week, sparking a 7 percent\\rise in its shares.                                                             \\ \hline
(IGA) Prediction =   \textbf{Business}. (Confidence = \textbf{0.920} )                                                                                                                                                                                                                                                                               \\ \hline
TheStreet.{\color{red}kom} May Be Up for Sale --   Report (Reuters) Reuters - The\\Street.com Inc. , the financial {\color{red}novice} and   commentary {\color{red}Network} site,\\may be up for sale, according to a report in   Business Week, sparking\\a 7 percent rise in its shares.                   \\ \hline
(PSO$_{attack}$) Prediction =   \textbf{Business}. (Confidence = \textbf{0.983} )                                                                                                                                                                                                                                                                               \\ \hline
TheStreet.com May Be Up for Sale --   {\color{red}Exposition} (Reuters) Reuters -\\TheStreet.com Inc. , the {\color{red}fiscal} news   and {\color{red}critique} Web {\color{red}locale}, may be up\\ for {\color{red}monopoly}, according to a report in Business Week, sparking a 7 per-\\cent rise in its {\color{red}stocks}. \\ \hline
(LEAP) Prediction =   \textbf{Business}. (Confidence = \textbf{0.998} )                                                                                                                                                                                                                                                                              \\ \hline
TheStreet.com May Be Up for Sale --   Report (Reuters) Reuters - The\\Street.com Inc. ,the financial news and   commentary {\color{red}vane} site, may be up\\for sale, according to a report in   {\color{red}job} Week, sparking a 7 percent rise in\\its shares.                                         \\ \hline
\end{tabular}}

\footnotesize{\vspace{0.5em} Note: As the text in AG's News is of moderate length, we use it to showcase the adversarial test cases. The modified words in the adversarial test cases are highlighted in red.}
 \vspace{-0.5cm}
\end{table}%
\begin{center}
\begin{tcolorbox}[colback=gray!10,
                  colframe=black,
                  width=8.5cm,
                  arc=1mm, auto outer arc,
                  boxrule=0.5pt,
                  left=1mm,
                  right=1mm,
                  top=1mm,
                  bottom=1mm,
                 ]
\textbf{Answer to RQ1:} LEAP generates higher-quality test cases for structurally different victim models and datasets with different characteristics, and it performs exceptionally well in terms of success rates. 
\end{tcolorbox}
\end{center}

\vspace*{0.2em}\noindent\textbf{RQ2: Can LEAP generate test cases more efficiently?}\vspace*{0.2em}

\begin{figure}[t]
 \centering
\includegraphics[width=0.95\linewidth]{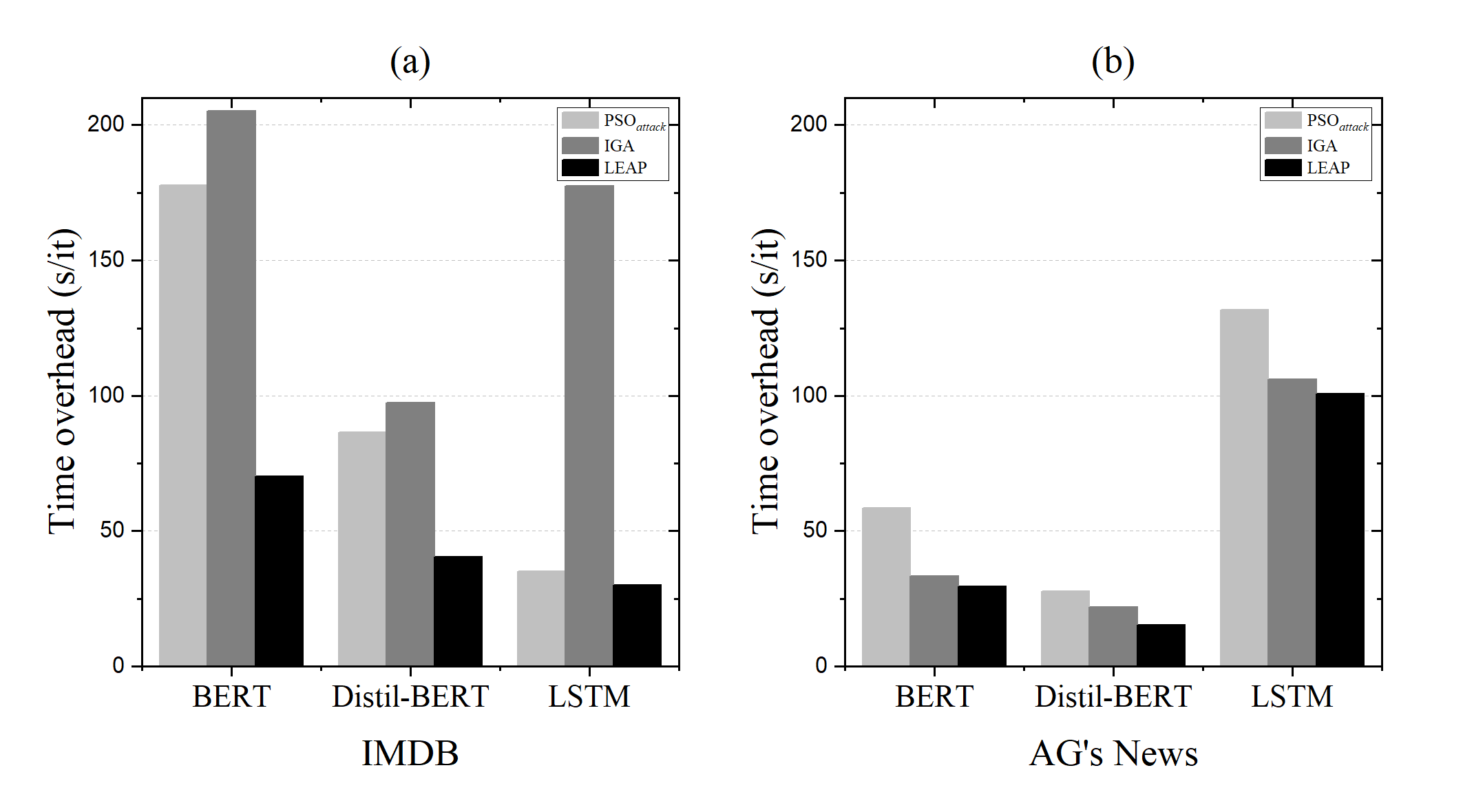}\\
 \caption{Results of the time overhead for testing different victim models. The lower the values are, the more efficient the method is.}\label{Fig4}
 \vspace{-1cm}
\end{figure}

Apart from the quality of test cases, the efficiency of the testing method, including time overhead and query number, is also our main concern. Fig. \ref{Fig4} shows the time overhead of generating test cases for the long text datasets IMDB and AG's News. As we can see, for all victim models, LEAP has less time overhead per successfully generated test case. On average, LEAP is  2.14s\textasciitilde147.57s faster than the best baseline per generated test case. 
When testing BERT finetuned by AG's News, the time overheads of IGA, PSO$_{attack}$, and LEAP are 205.28s/it, 177.81s/it, and 70.17s/it, which indicates that LEAP is more efficient. The vast majority of the testing process is spent on querying the victim models~\cite{9794089}, so the reduction in time overhead also indicates that LEAP has fewer query numbers. We show such results in the repository\footnote{https://github.com/lumos-xiao/LEAP} due to limited space.
\begin{center}
\begin{tcolorbox}[colback=gray!10,
                  colframe=black,
                  width=8.5cm,
                  arc=1mm, auto outer arc,
                  boxrule=0.5pt,
                  left=1mm,
                  right=1mm,
                  top=1mm,
                  bottom=1mm,
                 ]
\textbf{Answer to RQ2:} In terms of testing efficiency, LEAP can generate successful test cases with less time overhead and fewer query numbers, thus saving more testing time. 
\end{tcolorbox}
\end{center}

\vspace*{0.2em}\noindent\textbf{RQ3: How transferable are the test cases generated by LEAP?}\vspace*{0.2em}

\begin{figure}[t]
 \centering
\includegraphics[width=1.0\linewidth]{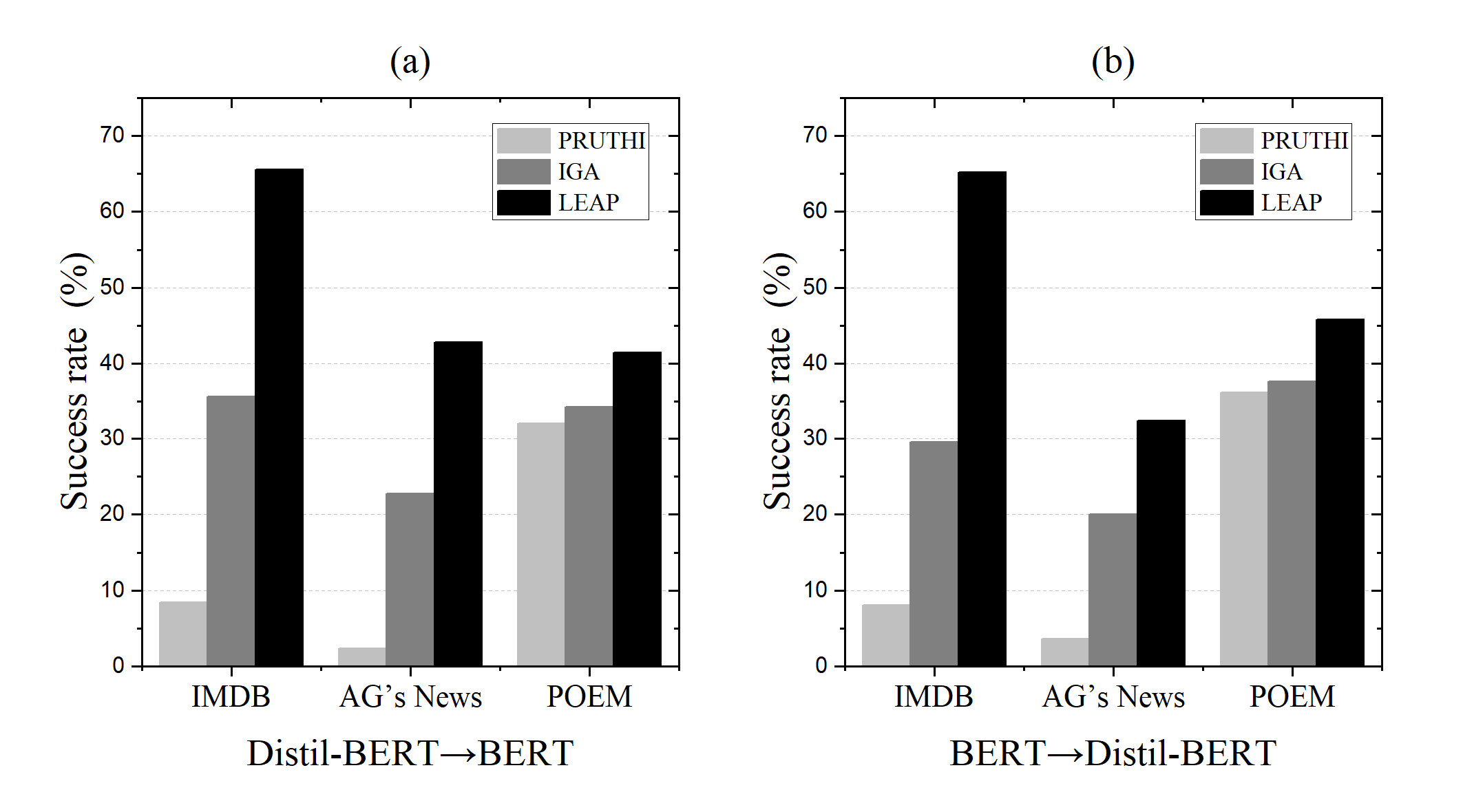}\\
 \caption{The success rates of transferred adversarial test cases on the three datasets (want $\uparrow$)}\label{Fig5}
 \vspace{-0.5cm}
\end{figure}

Figure \ref{Fig5} shows the transferability comparison of LEAP with the baselines, where we selected one baseline (i.e. IGA and PRUTHI) with excellent performance from heuristic and non-heuristic test methods, respectively. Fig. \ref{Fig5}(a) shows the success rate results of transferring the test cases made for testing Distil-BERT to BERT and vice versa in Fig. \ref{Fig5}(b). We find that, for the victim model finetuned on the three different types of datasets, the test cases generated by LEAP all exhibit the highest transferability, and the migrated test cases have a higher success rate~\cite{wang2020cat}. Taking the IMDB dataset as an example, the success rates of test cases generated by PRUTHI, IGA, and LEAP on BERT are 13.4\%, 90.8\%, and 92.2\%, respectively. The success rates of migration to Distil-BERT are 8.4\%, 35.6\%, and 65.6\%, and LEAP still maintains the highest success rate.
\begin{center}
\begin{tcolorbox}[colback=gray!10,
                  colframe=black,
                  width=8.5cm,
                  arc=1mm, auto outer arc,
                  boxrule=0.5pt,
                  left=1mm,
                  right=1mm,
                  top=1mm,
                  bottom=1mm,
                 ]
\textbf{Answer to RQ3:} Test cases generated by LEAP have higher transferability, which means that LEAP is able to uncover more defects in DNN-based systems even without access to their internal DNN models.
\end{tcolorbox}
\end{center}

\vspace*{0.2em}\noindent\textbf{RQ4: Whether the test cases generated by LEAP contribute to enhancing the robustness of the victim model?}\vspace*{0.2em}

For this research question, to simulate the low-resource scenario, we mixed the adversarial test cases generated from 10\% of the original training set with the original training set according to the experimental setting of \cite{li2021contextualized}. We used the IMDB with the most extended text length (i.e., 215 words/it) in our experimental datasets
as the original dataset, 
resulting in three adversarial training sets as shown in Table \ref{tab3}. The success rates of all the methods on different adversarial training datasets decreased, and the success rates on the victim models finetuned with IMDB$_{LEAP}$ are 3.9\%, 77.59\%, and 80.4\%, respectively, with the most significant decreases. This implies that the test cases generated by LEAP improve the model's robustness more than the other baselines, since a lower success rate demonstrates that the victim model correctly classifies more adversarial test cases. Notably, LEAP still manages to obtain the highest success rate regardless of which adversarial training set finetuned victim model is tested, which further illustrates the excellent performance of LEAP in mining the defects of DNN-based systems.
\begin{table}[]
\setlength{\tabcolsep}{3.5pt}
\centering
\caption{Performance comparison of test methods on BERT after adversarial training. }
\begin{tabular}{c|c|c|c|c|c}
\hline
\textbf{Baseline}                                                           & \textbf{Indicator} & \textbf{IMDB}  & \textbf{IMDB$_{PRU}$} & \textbf{IMDB$_{IGA}$} & \textbf{IMDB$_{LEAP}$} \\ \hline
\multirow{3}{*}{\textbf{\begin{tabular}[c]{@{}c@{}}PRU\\ THI\end{tabular}}} & \textit{S-rate}    & 0.134          & 0.081              & 0.076              & 0.039               \\
                                                                            & \textit{C-rate}    & \textbf{0.046} & \textbf{0.036}     & \textbf{0.037}     & \textbf{0.053}      \\
                                                                            & \textit{T-O(s/it)} & \textbf{5.164} & \textbf{4.052}     & \textbf{5.858}     & \textbf{4.599}      \\ \hline
\multirow{3}{*}{\textbf{IGA}}                                               & \textit{S-rate}    & 0.908          & 0.904              & 0.848              & 0.776               \\
                                                                            & \textit{C-rate}    & 0.123          & 0.137              & 0.162              & 0.166               \\
                                                                            & \textit{T-O(s/it)} & 33.195         & 26.387             & 58.124             & 69.357              \\ \hline
\multirow{3}{*}{\textbf{LEAP}}                                              & \textit{S-rate}    & \textbf{0.922} & \textbf{0.918}     & \textbf{0.909}     & \textbf{0.804}      \\
                                                                            & \textit{C-rate}    & 0.112          & 0.132              & 0.146              & 0.161               \\
                                                                            & \textit{T-O(s/it)} & 29.383         & 25.632             & 41.766             & 54.431              \\ \hline
\end{tabular}
\label{tab3}
 \vspace{-0.5cm}
\end{table}

We use the change rate to measure the quality of test cases. The adversarially trained victim models, especially those finetuned using IMDB$_{LEAP}$, force the test method  to increase the original text's perturbation to generate successful test cases. 
As shown in Table \ref{tab3}, when testing the victim model finetuned by IMDB$_{PRU}$ and IMDB$_{IGA}$, the change rate of PRUTHI becomes lower instead. We believe this is because PRUTHI increases the perturbation on the original text to generate mostly failed test cases, which leads to an excessive decrease in the success rate compared to the one on the original training set.
In addition, we observed that models finetuned by the adversarial training sets significantly increased the time overhead of the test methods, with the models finetuned using IMDB$_{LEAP}$ increasing the most. 
This also indicates that testing tools have the most difficulty to find successful adversarial test cases for the model finetuned with LEAP-generated test cases, which on the other hand indicates LEAP can improve the robustness of victim models.

\begin{center}
\begin{tcolorbox}[colback=gray!10,
                  colframe=black,
                  width=8.5cm,
                  arc=1mm, auto outer arc,
                  boxrule=0.5pt,
                  left=1mm,
                  right=1mm,
                  top=1mm,
                  bottom=1mm,
                 ]
\textbf{Answer to RQ4:} The training set with test cases generated by LEAP significantly reduces the success rate, case quality, and efficiency of the test methods. Therefore, the adversarial test cases generated by LEAP are efficacious for improving the robustness of the victim model. 
\end{tcolorbox}
\end{center}

\vspace*{0.2em}\noindent\textbf{RQ5: Does each of the method components proposed in this paper improve the quality of the generated test cases and the testing efficiency of LEAP?}\vspace*{0.2em}

We finetuned BERT as the victim model on three datasets, ablating each component of LEAP that is different from the most similar existing work PSO$_{attack}$ to investigate its effectiveness. Table~\ref{tab4} shows the experimental results using 1000 test examples. Since the test set of POEM only contains  104 examples, we sampled the test set 10 times using different random seeds with 100 examples each time.
On the IMDB dataset, LEAP only improves the success rate by 0.9\% compared to PSO$_{attack}$.
However, LEAP is nearly twice as fast as PSO$_{attack}$ in terms of time overhead. On AG's News, LEAP shows significant improvement in all the metrics. In particular, the use of Levy flight for population initialization and adaptive update operator increases the success rate by 11.6\%, reduces the change rate by 8.66\%, and decreases the time overhead by 107.64s. On POEM, the use of the greedy variation operator reduces the success rate of LEAP by 0.3\%, which is because the introduction of the greedy strategy increases the risk of the search algorithm falling into local optima in high-dimensional text data. However, it effectively reduces the change rate by 0.8\% and the time overhead by 0.41s. Overall, despite the datasets being from different domains with different textual features, LEAP's improved strategy achieves better test results than PSO$_{attack}$.
\begin{table}[]
\centering
\caption{Results of ablation study on Levy flight-based population initialization, Adaptive update particles (\emph{adaptive}), and greedy mutation (\emph{greedy}).}
\begin{tabular}{c|c|c|c|c}
\hline
\textbf{Dataset}                    & \textbf{Testing method}   & \textbf{S-rate} & \textbf{C-rate} & \textbf{T-O(s/it)} \\ \hline
\multirow{4}{*}{\textbf{IMDB}}      & PSO$_{attack}$                          & 0.913           & 0.173           & 58.533             \\
                                    & \textit{w/o adaptive,greedy} & 0.916           & 0.135           & 44.026             \\
                                    & \textit{w/o greedy}          & 0.916           & 0.118           & 34.785             \\
                                    & LEAP                         & \textbf{0.922}  & \textbf{0.113}  & \textbf{29.380}    \\ \hline
\multirow{4}{*}{\textbf{AG's News}} & PSO$_{attack}$                          & 0.696           & 0.244           & 177.811            \\
                                    & \textit{w/o adaptive,greedy} & 0.712           & 0.178           & 123.042            \\
                                    & \textit{w/o greedy}          & 0.778           & 0.158           & 99.688             \\
                                    & LEAP                         & \textbf{0.812}  & \textbf{0.157}  & \textbf{70.174}    \\ \hline
\multirow{4}{*}{\textbf{POEM}}      & PSO$_{attack}$                          & 0.658           & 0.196           & 1.457              \\
                                    & \textit{w/o adaptive,greedy} & 0.690           & 0.189           & 1.930              \\
                                    & \textit{w/o greedy}          & 0.711           & 0.169           & 1.835              \\
                                    & LEAP                         & \textbf{0.714}  & \textbf{0.161}  & \textbf{1.426}     \\ \hline
\end{tabular}
\label{tab4}
\end{table}
\begin{center}
\begin{tcolorbox}[colback=gray!10,
                  colframe=black,
                  width=8.5cm,
                  arc=1mm, auto outer arc,
                  boxrule=0.5pt,
                  left=1mm,
                  right=1mm,
                  top=1mm,
                  bottom=1mm,
                 ]
\textbf{Answer to RQ5:} Compared to the most similar existing work, LEAP's components are effective in generating high-quality test cases more efficiently. 
\end{tcolorbox}
\end{center}

\section{Threats to Validity}
Our experimental results demonstrate LEAP’s effectiveness. However, we also acknowledge some threats to validity.

\textbf{Internal validity}. The main threat comes from the setting of hyperparameters in the experiments, such as population size and the maximum number of iterations. To mitigate the threat, we use the same hyperparameters for all experiments on each dataset, and try to choose the same hyperparameters as the existing method PSO$_{attack}$ to show the validity of our method.

\textbf{External validity}. Our experiments focused on testing DNNs in an English environment, which may threaten the generality of LEAP for other languages. But applying LEAP on DNNs in other languages requires only minor input adjustments. We mitigate this threat by evaluating our approach on three kinds of datasets and three types of NLP models. This makes us confident that LEAP will work across a variety of NLP applications.
\section{Related Work}

\textbf{Testing AI Software.} The development of Artificial Intelligence (AI) software has been gaining momentum in recent years, with a growing need for effective testing strategies to ensure their reliability and performance. Automated testing techniques have been widely used by software professionals due to their efficiency, cost-effectiveness and reusability. In the field of Computer Vision (CV), a large number of automated testing techniques have been proposed~\cite{yuan2021meta, rony2021augmented,zhang2021cagfuzz}. 
The primary difference between NLP and CV software is that the feature space of text data is discrete, and any modifications to the original example are more likely to result in errors in semantics and sentence fluency, which can be easily detected~\cite{liang2018deep}. Morris et al.~\cite{morris2020textattack2} decomposed the testing process into four components: goal function, constraint list, transformation, and search method, and unified them within the Python framework TextAttack. Tan et al.~\cite{tan2021reliability} demonstrated the incorporation of adversarial attacks as reliability tests into the reliability testing framework DOCTOR, presenting a method to enhance accountability in existing efforts.
Overall, there are three main types of DNN-based automated testing methods for AI software: 1) white-box testing methods~\cite{ebrahimi2018hotflip,yoo2021towards2} based on internal information such as DNN gradients, 2) greedy methods~\cite{2021Beyond,pruthi2019combating} that modify the text at each specific index to minimize the original DNN prediction, and 3) heuristic methods~\cite{wang2019natural,zang-etal-2020-word} that heuristically search for the optimal option among potential test cases. 

\textbf{Testing NLP Software.}
In the field of NLP, Ribeiro et al.~\cite{ribeiro2022adaptive} utilized large-scale language models and human feedback to generate adaptive unit tests for victim models. Wu et al.~\cite{wu2019errudite} developed Errudite, an interactive tool that utilizes domain-specific language to facilitate precise error grouping and analysis. In contrast, our paper focuses on automating the testing of NLP software, taking into account time and cost constraints. Based on the minimum perturbation units used in applications, related works are divided into three aspects:

\textbf{1) Sentence-level method.}
Sentence-level testing methods are more flexible in terms of perturbation, and the modified sentence can be inserted in any part of the text when the semantics and syntax are correct. It is executed by adding ordered words of a certain length.
Sentence-level methods are widely used in Q\&A~\cite{gan2019improving,wallace2019trick} and machine understanding~\cite{lin2021using,wang2018robust} systems, but have yet received more research in text classification~\cite{xu2021grey}. 
Since the sentence-level method modifies the entire sentence with a substantial impact on the semantics of the paragraph~\cite{
10.1145/3533767.3534394}, the naturalness of the generated test cases is particularly affected. Even if the test is successful, it is often incomprehensible to humans. In contrast, our method only modifies individual words of the original text with controlled modification restrictions, thus ensuring better naturalness.

\textbf{2) Char-level method.}
Char-level methods aim to modify a few characters within a word to generate test cases that cause DNNs to make decisions incorrectly~\cite{eger2019text,ebrahimi2018adversarial}. Given that the modifications typically involve spelling errors, Li et al.~\cite{li2018textbugger} generated adversarial test cases by inserting, swapping, and deleting specific characters, combined with the Jacobian matrix of the victim model.
Since character-level methods are prone to produce misspelled words~\cite{yang2021besa}, today's splendid spell-checking tools can easily detect such perturbations. In contrast, LEAP plans the perturbation space utilizing a lexical network to produce a synonym dictionary, and the potential perturbations are all actual words, so there is no problem with misspellings.

\textbf{3) Word-level method.}
Word-level methods perturb text by inserting, deleting, and replacing whole words, which is significantly better than other methods in naturalness and transferability, and therefore has gained the most attention~\cite{lee2022query,yuan2021transferability}.
Li et al.~\cite{li2020bert} utilized pre-trained language models as masked models to generate substitute words, considering contextual information. Jin et al.~\cite{jin2020bert} employed word importance ranking and cosine similarity between word vectors for synonym replacement. Ye et al.~\cite{ye2022texthoaxer} formulated a hard-label scenery as an optimization problem based on gradient perturbation metrics in word embedding space, generating test cases with smaller query budgets and higher semantic similarity.
LEAP is a word-level testing method that uses PSO to determine the words to be replaced and redesigns the internal arithmetic of PSO by combining the features of NLP test cases. This allows our method to guarantee the same high success rate as other heuristic testing methods while requiring less time overhead and fewer queries.
\section{Conclusion}
In this paper, we propose LEAP, a black-box testing method for DNN-based NLP systems that efficiently generates adversarial test cases. To address the problems of low data utilization and high time overhead in current testing methods, we design new components for discrete text data, including initializing populations using Levy flight, adaptively updating particles, and employing a greedy mutation approach. We evaluate the performance of LEAP using three datasets, three advanced deep learning models, and five baselines. The experimental results demonstrate that the average success rate of adversarial test cases generated by LEAP is 79.1\%, surpassing other baselines, and that the time overhead is reduced by 2.14s\textasciitilde147.57s compared to other heuristic-based methods. 
We also investigate the value of adversarial test cases generated by LEAP in enhancing the robustness of victim models.

For future work, we plan to enhance the scalability of LEAP to encompass a broader range of NLP downstream tasks and accommodate more complex perturbation scenarios, including character-level or sentence-level modifications. To achieve this, we will explore modular granularity settings and adaptive search algorithms as potential solutions.

\section*{Acknowledgment}
This work is supported by the National Natural Science Foundation of China under Grants 62272145 and U21B2016.

\bibliographystyle{ieeetr}
\bibliography{ref}

\end{document}